\documentclass{ws-procs975x65}            

\newcommand{\be}{\begin{equation}}
\newcommand{\ee}{\end{equation}}
\newcommand{\bea}{\begin{eqnarray}}
\newcommand{\eea}{\end{eqnarray}}

\begin{document}
\title{Adiabatic regularization with a Yukawa interaction}

\author{Antonio Ferreiro}
\address{Departamento de Fisica Teorica and IFIC, Universidad de Valencia-CSIC,\\
Facultad de Fisica, Burjassot, 46100, Spain\\
$^*$E-mail: antonio.ferreiro@ific.uv.es}

\author{Adrian del Rio} 

\address{Centro de Astrof\'isica e Gravita\c{c}\~ao - CENTRA,\\
Departamento de F\'isica, Instituto Superior T\'ecnico - IST,\\
Universidade de Lisboa - 1049 Lisboa, Portugal}

\author{Jose Navarro-Salas and Silvia Pla}
\address{Departamento de Fisica Teorica and IFIC, Universidad de Valencia-CSIC,\\
Facultad de Fisica, Burjassot, 46100, Spain}

\author{Francisco Torrenti} 
\address{Department of Physics, University of Basel,\\ Klingelbergstr.~82, CH-4056 Basel, Switzerland}

\begin{abstract}

We extend the adiabatic regularization method for an expanding universe to include the Yukawa interaction between a quantized Dirac field  and a homogeneous time-dependent scalar field. We present the renormalized semiclassical equations that are needed in order to take into account the backreaction of the produced Dirac fermions in both gravitational and scalar background fields. 
\end{abstract}

\bodymatter

\section{Introduction}
 
Inflation typically predicts an exponentially fast dilution of the number density of any particle species. To repopulate the Universe with matter and radiation, the energy density of the inflationary sector must be transferred to standard model degrees of freedom. This epoch is called reheating. An initial preheating stage, where particles are created due to the non-adiabatic time-dependence of the inflaton field, has been proposed for both scalar \cite{spreheating} and fermion fields\cite{fpreheating}. For example, in models such as chaotic inflation, the inflaton oscillates around the minimum of its potential after inflation, and this excites particles species coupled to it (both bosons and fermions) due to resonant effects.

 Gravitational particle creation was first discovered in the expanding universe\cite{parker68, parker-toms, birrell-davies}  and later extended to black holes\cite{hawking74, parker-toms, birrell-davies, Fabbri-Navarro}.  Time-dependent gauge fields can also create particles\cite{BI}. Due to the Yukawa coupling between a time-varying classical scalar field $\Phi$ and a quantum fermionic field $\psi$, spontaneous particle production also occurs. In order to take into account the backreaction of the produced particles on the classical background fields, via the semiclassical equations, one need to compute the expectation values $\langle T_{\mu\nu} \rangle$ and $\langle \bar \psi \psi\rangle$. This requires a regularization and renormalization method to deal with the ultraviolet divergences appearing in expectation values quadratic in fields. 
 A very powerful method to tame the ultraviolet divergences for isotropically expanding universes was originally proposed by Parker and Fulling\cite{parker-fulling, parker-toms, birrell-davies} for scalar fields. This renormalization method is based on an adiabatic expansion of the field modes living  in the expanding universe, and it is known as adiabatic regularization. The adiabatic method has been extended to Dirac fermions in\cite{LNT, BFNV}.

In this contribution we further extend the adiabatic regularization method  to incorporate the Yukawa interaction, which is expected to be  of major importance  in the preheating epoch.  First, we present the semiclassical equations for fermions in an expanding universe and Yukawa-coupled to a scalar field and then use the adiabatic regularization scheme  to cure the divergences of the vacuum expectation values of  $\langle T_{\mu\nu} \rangle$ and $\langle \bar \psi \psi\rangle$. (For a extended analysis, see \cite{DFNT}). 

\section{Semiclassical equations for a quantized Dirac matter field with Yukawa coupling}
We consider the theory defined by the action functional  $S=S[ g_{\mu\nu},\Phi,\psi,\nabla\psi]$,
where $\psi$ represents a  Dirac field, $\Phi$ is a scalar field, and $g_{\mu\nu}$ stands for the  spacetime metric. We decompose the action as $S=S_{g}+S_m$, where $S_m$ is the matter sector
\be \label{Sm} S_m = \int d^4x   \sqrt{-g} \  \left\{ \frac{i}{2}[ \bar \psi \underline\gamma^{\mu} \nabla_{\mu} \psi -(\nabla_{\mu} \bar \psi)\underline\gamma^{\mu} \psi)] - m\bar \psi \psi -g_Y \Phi \bar \psi \psi \right\} \ ,  \ee
and $S_g$ is the gravity-scalar sector. In (\ref{Sm}), both the metric $g_{\mu\nu} (x)$ and the scalar  field $\Phi (x)$ are regarded as classical external fields. The curved-space Dirac matrices $\underline\gamma^{\mu}$ obey the relations $\{ \underline\gamma^{\mu}, \underline\gamma^{\nu} \} = 2 g^{\mu\nu}$. The Dirac spinor $\psi (x)$ will be our quantized field, living in a curved spacetime and possessing  a Yukawa coupling to the classical  field $\Phi$. The Dirac equation is 
 \be (i  \underline\gamma^{\mu}\nabla_\mu -m - g_Y \Phi) \psi =0 \ , \ee
and the stress-energy tensor is given by
\be T_{\mu\nu}^m :=\frac{2}{\sqrt{-g}}\frac{\delta S_m}{\delta g^{\mu\nu}}  
 =  \frac{i}{2} \left[\bar \psi \underline\gamma_{(\mu}\nabla_{\nu)}\psi - (\nabla_{(\mu}\bar \psi)\underline\gamma_{\nu)} \psi \right]  \label{EMtensor}\ .  \ee

The complete theory, including the gravity-scalar sector in the action, can be described by
\be S=  S_g + S_m = \frac{1}{16\pi G}\int d^4x\sqrt{-g} R + \int d^4x\sqrt{-g} \left\{ \frac{1}{2}g^{\mu\nu}\nabla_\mu \Phi \nabla_\nu \Phi
 - V(\Phi) \right\} + S_m \ ,  \label{eq:full-action}\ee
where $S_m$ is the action for the matter sector given in (\ref{Sm}). 

The semiclassical Einstein and scalar equations are
\bea G^{\mu\nu} +8\pi G (\nabla^\mu\Phi \nabla^\nu\Phi - \frac{1}{2}g^{\mu\nu} \nabla^\rho\Phi \nabla_\rho\Phi + g^{\mu\nu} V(\Phi)) &=& -8\pi G \langle T_m^{\mu\nu} \rangle_{ren} \ , \label{eqg2b} \\*
\Box \Phi + \frac{\partial V}{\partial \Phi} &=& - g_Y \langle \bar \psi \psi \rangle_{ren}\ \label{eqPhi2}. \eea
In order to obtain the renormalized quantities $\langle T_{\mu\nu} \rangle_{ren}$ and $\langle \bar \psi \psi\rangle_{ren}$ we extend the adiabatic regularization in the next section.

\section{Fermion quantization and adiabatic regularization}

In a spatially flat FLRW spacetime, we expand the field $\psi$ as 
\bea
\psi(x)=\int d^3\vec{k} \sum_{\lambda}\left[B_{\vec{k} \lambda}u_{\vec{k} \lambda}(x)+D_{\vec{k} \lambda}^{\dagger}v_{\vec{k} \lambda}(x) \right] \ ,  \label{4c}
\eea 
with 
\bea \label{uk}
&u_{\vec{k},\lambda}(x)= \frac{e^{i\vec k \vec x}}{\sqrt{(2\pi)^3a^{3} (t)}}
\left( {\begin{array}{c}
  h^I_{{k}}(t) \xi_{\lambda} (\vec{k}) \\
  h^{II}_{{k}}(t)\frac{\vec{\sigma}\vec{k}}{k} \xi_{\lambda} (\vec{k})\\
 \end{array} } \right) \ 
\eea
and $v_{\vec{k},\lambda}(x)=C u_{\vec{k},\lambda}(x)$. The modes  $v_{\vec{k},\lambda}(x)$ can be obtained by applying a charge conjugate transformation $C\psi = -i\gamma^2\psi^*$. Here  $\xi_{\lambda}$ with $\lambda ={\pm} 1$ are two constant orthonormal two-spinors ($\xi_{\lambda}^{\dagger}\xi_{\lambda'}=\delta_{\lambda,\lambda'}$), eigenvectors of the helicity operator $\frac{\vec{\sigma} \vec{k}}{2 k} \xi_{\lambda} = \frac{\lambda}{2} \xi_{\lambda}$. The time-dependent functions $h_k^{I}$ and $h_k^{II}$ satisfy the first-order coupled equations
\be  h_k^{II} = \frac{i a }{k} \left( \frac{\partial h_k^I}{\partial t} + i (m + g_Y \Phi)h_k^I \right) \ , \,\,\,\,\,\,\,\,\,\, h_k^{I} =  \frac{i a }{k} \left( \frac{\partial h_k^{II}}{\partial t} - i (m + g_Y \Phi)h_k^{II} \right) \label{ferm-hk2b} \ . \ee
The normalization condition for the above four-spinors reduces to 
\be |h_k^{I}|^2 + |h_k^{II}|^2=1 \ . \label{ferm-wronsk2} \ee 

For adiabatic regularization one can expand  $h_k^I$ and $h_k^{II}$ as
  \bea h_k^{I} (t) = \sqrt{\frac{\omega(t) + m}{2 \omega(t)}} e^{-i \int^t \left( \omega(t') + \omega^{(1)}(t')+\dots\right) dt'} \left(1 + F^{(1)} (t)+\dots\right)\nonumber \\ h_k^{II} (t) = \sqrt{\frac{\omega(t) - m}{2 \omega(t)   }} e^{-i \int^t \left( \omega(t') + \omega^{(1)}(t')+\dots\right)  dt'}\left( 1 + G^{(1)}(t)+ \dots \right) \ . \label{ferm-ansatz} \eea 
  Here, $F^{(n)}$, $G^{(n)}$ and $\omega^{(n)}$ are complex and real functions of $n$th adiabatic order. By substituting (\ref{ferm-ansatz}) into the equations of motion (\ref{ferm-hk2b}) and the normalization condition (\ref{ferm-wronsk2}), one can obtain expressions for this functions by solving order by order in the adiabatic expansion. As usual, we consider $\dot a$ of adiabatic order 1, $\ddot a$ of adiabatic order 2, and so on. On the other hand, we consider the interaction term $s(t)$ of adiabatic order 1. For a recent discussion on a related issue see \cite{FNP}. We work out the first adiabatic order but the full computation until fourth adiabatic order can be found in \cite{DFNT, BFNV}. 
   In order to compute the three functions of the first adiabatic order contributions $F^{(1)}$,  $G^{(1)}$ and $\omega^{(1)}$ we treat independently the real and imaginary parts by writing $F^{(1)} = f_x^{(1)} + i f_y^{(1)}$ and $G^{(1)} = g_x^{(1)} + i g_y^{(1)}$. From  (\ref{ferm-ansatz}) and  (\ref{ferm-wronsk2}) we obtain for the real part
\bea
(\omega - m) (g_x^{(1)} - f_x^{(1)} ) &=& \omega^{(1)} - g_Y \Phi \ , \nonumber \\*
(\omega + m) (g_x^{(1)} - f_x^{(1)} ) &=& - \omega^{(1)} - g_Y \Phi \ , \nonumber \\*
(\omega + m) f_x^{(1)} + (\omega - m) g_x^{(1)} &=& 0 \ ,
\eea
which has as solutions
\be f^{(1)}_x = \frac{g_Y \Phi}{2 \omega} - \frac{m g_Y \Phi}{2 \omega^2} \ , \,\,\,\,\,\,\,\,\,\,\,\, g^{(1)}_x = - \frac{g_Y \Phi}{2 \omega} - \frac{m g_Y \Phi}{2 \omega^2} \ , \,\,\,\,\,\,\,\,\,\,\,\, \omega^{(1)} = \frac{m g_Y \Phi}{\omega} \ .\ee
On the other hand, the imaginary part of the system gives
\bea
(\omega - m) (g_y^{(1)} - f_y^{(1)} ) &=& \frac{1}{2} \frac{d \omega}{dt} \left( \frac{1}{\omega + m} - \frac{1}{\omega} \right) \ , \nonumber \\*
(\omega + m) (g_y^{(1)} - f_y^{(1)} ) &=& - \frac{1}{2} \frac{d \omega}{dt} \left( \frac{1}{\omega - m} - \frac{1}{\omega} \right) \ .
\eea
These two equations are not independent. The obtained solution for $g_y^{(1)}$ and $f_y^{(1)}$ is
\be f^{(1)}_y = A - \frac{m \dot a}{2 a \omega^2} \ , \,\,\,\,\,\,\,\,\,\,\,\, g^{(1)}_y = A \ , \ee
where $A$ is an arbitrary first-order adiabatic function. 
We will choose the simplest solution 
\be f^{(1)}_y = - \frac{m \dot{a}}{4 \omega^2 a} \ , \,\,\,\,\,\,\,\,\,\,\,\,  g^{(1)}_y = \frac{m \dot{a}}{4 \omega^2 a} \ ,\ee
obeying the condition
 $F^{(1)} (m, g_Y \Phi) = G^{(1)} (-m, -g_Y \Phi)$. Therefore, the adiabatic expansion will also preserve the symmetries of the equations (\ref{ferm-hk2b}) with respect to the change $(m,g_Y \Phi) \rightarrow (-m,-g_Y \Phi)$. We have checked that physical expectation values are independent to any potential ambiguity in this kind of  choice. We want to remark that this ambiguity  is originated by the possibility of adding  a time derivative in the integral of the exponent in (\ref{ferm-ansatz}) and compensating it by redefining the corresponding functions $F$ and $G$. Therefore, the local subtraction terms for renormalization can be unambiguously fixed. 
Nevertheless, an unambiguous adiabatic expansion of the modes can be  constructed following the procedure described in Ref. \cite{BFNV}.
 The next adiabatic orders are obtained by iteration. 

\section{Renormalized backreaction equations}

The classical stress-energy tensor in a FLRW spacetime has  two independent components. For a Dirac field, they are  (no sum on $i$),
\be T_0^0 = \frac{i}{2} \left( \bar{\psi} {\gamma}^0 \frac{\partial \psi}{\partial t} - \frac{\partial \bar{\psi}}{\partial t} {\gamma}^0 \psi \right) \ , \hspace{0.5cm}
T_i^i  = \frac{i }{2a} \left( \bar{\psi} {\gamma}^i \frac{\partial \psi}{\partial x^i} - \frac{\partial \bar{\psi}}{\partial x^i} {\gamma}^i \psi \right) \ . \label{ft00-ftii} \ee

In the quantum theory, using \eqref{4c} the vacuum expectation values  of the stress-energy tensor take the form
\be
\left< T_{00}\right>=\frac{1}{2\pi^2 a^3}\int_{0}^{\infty} dk k^2 \rho_k (t)  \ , \hspace{0.5cm} \rho_k (t) \equiv 2 i \left( h_k^{I} \frac{\partial h_k^{I*}}{\partial t} + h_k^{II} \frac{\partial h_k^{II*}}{\partial t} \right) \ , \label{12}
\ee
and
\be \left< T_{ii}\right>=\frac{1}{2\pi^2 a}\int_{0}^{\infty} dk k^2 p_k (t)  \ , \hspace{0.5cm}  
p_k (t)\equiv-\frac{2k }{3a} ( h_k^{I} h_k^{II*}+h_k^{I*} h_k^{II} )  \ . \label{13}
\ee
On the other hand we have:
\bea
 \langle \bar{\psi} \psi\rangle=\frac{-1}{\pi^2a^3}\int_0^{\infty} dk k^2 \langle \bar{\psi} \psi \rangle_k     \ , \hspace{0.5cm} \langle \bar{\psi} \psi \rangle_k  \equiv   |h_k^{I}|^2 - |h_k^{II}|^2 \ . \label{eq:variance} \eea

In general, adiabatic renormalization proceeds by subtracting the adiabatic terms of the integrand that contains the ultraviolet divergences. For  $\langle \bar \psi \psi\rangle$ we need to subtract until third adiabatic order and for $\langle T_{\mu\nu} \rangle$ until fourth adiabatic order.
It is important to remark that  all contributions of a given adiabatic term of fixed (adiabatic) order must be taken into account in the subtraction, otherwise general covariance is not maintained. Also, one subtracts only the minimum number of terms required to get a finite result. For example, for the first adiabatic order for $ \langle \bar{\psi} \psi \rangle_k$ introducing \eqref{ferm-ansatz} in  \eqref{eq:variance} and selecting the corresponding first adiabatic order term of this expansion for the  the expectation values we obtain
\bea
 \langle \bar{\psi} \psi \rangle_k^{(1)} && =\frac{\omega+m}{2 \omega}\left(|F|^2\right)^{(1)}-\frac{\omega-m}{2\omega}\left(|G|^2\right)^{(1)} \nonumber \\ &&=\frac{\omega+m}{2 \omega}\left(2 f_x^{(1)}\right)-\frac{\omega-m}{2\omega}\left(2 g_x^{(1)}\right)= \frac{g_Y \Phi}{\omega }-\frac{m^2 g_Y \Phi}{\omega ^3}  \label{eqbilineal2}\ .
\eea

The complete renormalized expressions for the energy momentum tensor and $\langle \bar \psi \psi\rangle_{ren}$ are
\bea \langle T_{00} \rangle_{ren} \equiv   \langle T_{00} \rangle - \langle T_{00} \rangle_{Ad}=      \frac{1}{2 \pi^2 a^3} \int_0^{\infty} dk k^2 ( \rho_k - \rho_k^{(0-4)}) \ , \label{ren-ten} \\
\left< T_{ii}\right>_{ren}\equiv \left< T_{ii}\right> - \left< T_{ii}\right>_{Ad}= \frac{1}{2\pi^2 a}\int_{0}^{\infty} dk k^2 (p_k-p_k^{(0-4)}) \ ,  \label{17b}
\\
 \langle \bar{\psi} \psi\rangle_{ren}=\langle \bar{\psi} \psi\rangle - \langle \bar{\psi} \psi\rangle_{Ad}=  \frac{-1}{\pi^2a^3}\int_0^{\infty} dk k^2 (\langle \bar{\psi} \psi\rangle_k-\langle \bar{\psi} \psi\rangle_k^{(0-3)})  \label{eq:ren-variance} \ . \eea 
 where $^{(0-n)}$ denotes the substraction terms of adiabatic order $0, 1, ..., n$. 

These finite renormalized quantities are the ones that are introduced in the semiclassical equations of 
\eqref{eqg2b} and \eqref{eqPhi2} in order to take into account the backreaction of the produced fields in both the gravitational and the scalar background fields.

We want also to briefly recall that the adiabatic method serves not only as a regulator but also as a renormalization mechanism. Each adiabatic subtraction of $\langle \bar \psi \psi\rangle_{ren}$ and  $\langle T_{\mu\nu} \rangle_{ren}$ can be understood as  the contribution  of a generic Lagrangian containing all  possible counterterms having couplings with non-negative mass dimension, up to Newtons's coupling constant \cite{DFNT}. The associated running of the couplings due to this adiabatic renormalization scheme has also been recently studied in \cite{FN} for a Maxwell-Einstein theory. This is still an open problem for a Yukawa interaction.  
Another issue is the correct fixing of the initial conditions in order to perform a numerical analysis of the full renormalized semiclassical equations. This has already been done for a scalar field in expanding universe \cite{agullo} but a  generalization for Dirac fields and also for Yukawa couplings is still missing 	\cite{FNPT}. 

\section{Conclusions}

In order to take into account the backreaction of the produced Dirac particles during the preheating stage one needs to compute  $\langle T_{\mu\nu} \rangle$ and $\langle \bar \psi \psi\rangle$, which are  plagued with UV divergences. To renormalize these observables we have extended the adiabatic regularization  to include a Yukawa interaction between the quantized Dirac fields and the classical scalar field background.

\end{document}